\documentstyle[preprint,aps,epsf]{revtex}

\newcommand{\ordnung}{\mathop{  \rm O  }\nolimits}
\begin{document}
\draft
\bibliographystyle{revtex}
\title{Testing the time dependence of the fundamental constants
       in the spectra of multicharged ions}
\author{O.~Yu.~Andreev,${}^1$ L.~N.~Labzowsky,${}^{2,3}$
        G.~Plunien,${}^1$ and G.~Soff${}^1$}
\address{${}^1$ {Institut f\"ur Theoretische Physik,
             Technische Universit\"at Dresden,
             Mommsenstra{\ss}e 13, D-01062, Dresden, Germany}}
\address{${}^2$ {V.~A.~Fock~Institute of Physics,
             St.~Petersburg State University,
             Ulyanovskaya 1,
             198504,
             St.~Petersburg, Russia}}
\address{${}^3$ {Petersburg Nuclear Physics Institute,
             188300,
             Gatchina, St.~Petersburg, Russia}}
\date{\today}
\maketitle
\begin{abstract}
A new method for measuring a possible time dependence of the fine-structure constant ($\alpha$) is proposed.
The method is based on the level-crossing in two-electron
highly-charged ions facilitating resonance laser measurements of
the
distance between the levels at
the point of crossing.
This provides an enhancement factor of about
$10^{3}$
in Helium-like Europium
and
thus reduces
the requirements for the relative accuracy of
resonance laser measurements at about
$10^{-12}$.
\end{abstract}
\pacs{PACS number(s): 31.30.Jv, 31.10.+z}
Questions about the constancy of fundamental
coupling and mass parameters in
physics, first addressed by Dirac
\cite{dirac37},
now become of particular interest in connection with new
developments towards unified field theories, such as string
theories, D-brane models,
etc.
These theories, in principle, allow for a space and time
dependence of
fundamental constants (see
Ref.~\cite{uzan03}
for a comprehensive review on the subject).
Presently, the situation regarding this problem appears to be very stringent, since spectroscopical studies of the spectra of quasars
indicated a deviation of the fine-structure constant from its
standard value of about
${\Delta \alpha} / {\alpha}=(-0.76\pm 0.28)\times 10^{-5}$
for values of the redshift parameter $z$
within the range
$z=1.8-3.5$
\cite{webb01}.
This deviation is related to the logarithmic time derivative
$\dot{\alpha} / {\alpha} =
(-0.38 \pm 0.14)\times 10^{-14} \,\mbox{\rm yr}^{-1}$
and suggests that
$\alpha$
may have been smaller in the past.
From the other side, the results
obtained from quasar data
strongly contradict with those deduced
from the natural fission reactor in Oklo, which yield
$\dot{\alpha} / {\alpha} =
(-0.2 \pm 0.8)\times10^{-17} \,\mbox{\rm yr}^{-1}$
\cite{fujii02,uzan03}.
This contradiction
may be also interpreted as indication for a spacial variation of
$\alpha$.
Accordingly, laboratory atomic measurements are highly necessary for clarifying the situation.

\par
The best empirical bounds are set by the comparison of
the two-photon
$1s-2s$
resonance frequency in hydrogen to a Cs clock during 4 years
which yield the result
$\dot{\alpha} / \alpha = (-0.9 \pm 2.9)\times 10^{-15}
\,\mbox{\rm yr}^{-1}$
\cite{fischer04},
and by the comparison of an optical transition in
Yb${}^{+}$
ion to Cs clock during 2.8 years
$\dot{\alpha} / \alpha = (-0.3 \pm 2.0)\times 10^{-15}
\,\mbox{\rm yr}^{-1}$
\cite{peik04}.

\par
Within this paper we propose a new type of atomic experiments
for the search for variations
of the fine-structure constant
employing the crossings of atomic levels as a function of
the nuclear charge number
$Z$
in two-electron highly-charged ions (HCI).
These energy levels including QED
corrections were recently evaluated with high accuracy in
\cite{andreev03}.
The effect of the possible time-variation of $\alpha$
becomes strongly enhanced when it is investigated close to such
crossing points.

\par
We consider the energy difference
$\Delta E$
between two energy levels in Helium-like HCI as a function of
the two parameters
$\alpha$ and $Z$, respectively.
Accurate calculation of the energy levels within
the framework of QED should account for
a number of energy corrections
corresponding to various effects.
In terms of these corrections the function
$\Delta E$
can be represented as the sum
\begin{eqnarray}
\Delta E (\alpha,Z)
&=&\label{deltaEf}
  \Delta E^{\text{NS}} (Z)
+\Delta E^{\text{1ph}} (\alpha, Z)
+\Delta E^{\text{SE+VP}} (\alpha, Z)
+\Delta E^{\text{2ph}} (\alpha, Z)
+\Delta E^{\text{Scr(SE+VP)}} (\alpha, Z)
\,,
\end{eqnarray}
where the nuclear-size correction is of order
$\alpha^{0}$.
In the following, we shall focus on the two-electron levels
$(1s,2s)$
and
$(1s,2p)$,
respectively.
In case of a point nucleus the Dirac energies of
$2s$- and $2p$-states are equal, correspondingly,
finite-nuclear size induces a shift
$\Delta E^{\text{NS}}$
which is supposed to be independent on
$\alpha$.
The contributions
$\Delta E^{\text{1ph}}$
and
$\Delta E^{\text{2ph}}$
refer to one- and two-photon exchange corrections, respectively.
The term
$\Delta E^{\text{SE+VP}}$
represents the sum of the first-order (one-electron)
self-energy (SE)
and vacuum-polarization (VP) corrections.
Finally, the term
$\Delta E^{\text{Scr(SE+VP)}}$
denotes the corresponding screening corrections.
Each individual energy shift can be decomposed
into a power series with respect to
$\alpha$
and $Z$
with coefficients
$C^{\text{correction}}_{\text{powers of} \,\alpha, Z}$.
Accordingly, the one-photon exchange correction is
taken into account
up to terms
\begin{eqnarray}
\Delta E^{\text{1ph}} (\alpha, Z)
&=&\label{deltaEf1ph}
  C^{\text{1ph}}_{0,1} Z
+ C^{\text{1ph}}_{2,3}\alpha^2 Z^3
+ C^{\text{1ph}}_{4,5}\alpha^4 Z^5
\,.
\end{eqnarray}
The one-loop radiative effects are proportional to
$Z(\alpha Z)^3$
\begin{eqnarray}
\Delta E^{\text{SE+VP}} (\alpha, Z)
&=&\label{deltaEfsevp}
  C^{\text{SE+VP}}_{3,4} \alpha^3 Z^4
\,.
\end{eqnarray}
In comparison with
$\Delta E^{\text{1ph}}$
the two-photon exchange correction
$\Delta E^{\text{2ph}}$
is smaller by one factor of $Z$.
Accordingly, it can be expanded up to terms
$\ordnung (\alpha^5 Z^5)$
\begin{eqnarray}
\Delta E^{\text{2ph}} (\alpha, Z)
&=&\label{deltaEf2ph}
  C^{\text{2ph}}_{0,0}
+ C^{\text{2ph}}_{2,2}\alpha^2 Z^2
+ C^{\text{2ph}}_{4,4}\alpha^4 Z^4
\,.
\end{eqnarray}
The last term in
Eq.~(\ref{deltaEf})
is a two-photon correction which accounts for
the effect of screening of SE and VP
\begin{eqnarray}
\Delta E^{\text{Scr(SE+VP)}} (\alpha, Z)
&=&\label{deltaEfscrsevp}
  C^{\text{Scr(SE+VP)}}_{3,3} \alpha^3 Z^3
\,.
\end{eqnarray}

\par
While
the terms proportional to
$C^{\text{1ph}}_{0,1}$
and
$C^{\text{2ph}}_{0,0}$
correspond to
the pure Coulomb interaction between the two electrons,
the terms with coefficients
$C^{\text{1ph}}_{2,3}$
and
$C^{\text{2ph}}_{2,2}$
represent relativistic corrections
to the interelectronic interaction, in particular,
the Breit interaction.
The terms with coefficients
$C^{\text{1ph}}_{4,5}$
and
$C^{\text{2ph}}_{4,4}$
account for higher-order relativistic corrections
to the electron-electron interaction.

\par
Since we are interested in the relative change of
$\Delta E$
due to variations of the parameter $\alpha$,
one may consider the quantity
\begin{eqnarray}
\frac{\delta \Delta E(\alpha, Z)}{\Delta E (\alpha, Z)}
&=&\label{deltaEratio}
(\Delta E)^{-1}\,\,
\frac{\partial \Delta E(\alpha, Z)}{\partial \alpha}
\,\,
\delta \alpha
\,,
\end{eqnarray}
where
$\delta \alpha$
is the change of the fine-structure constant during the time interval
$\delta t$.
Introducing now
the relative variation
${\delta \alpha}/{\alpha} = ({\dot{\alpha} }/{\alpha})
\, \delta t$
yields
\begin{eqnarray}
\frac{\delta \Delta E(\alpha, Z)}{\Delta E (\alpha, Z)}
&=&\label{deltaEratio2}
\left[
(\Delta E)^{-1}\,\,
\alpha
\frac{\partial \Delta E(\alpha, Z)}{\partial \alpha}
\right]
\,
\frac{\dot{\alpha}}{\alpha}
\,
\delta t
\,.
\end{eqnarray}
An estimate for the logarithmic derivative
$\dot{\alpha} / \alpha$,
measured in the units of $\mbox{yr}^{-1}$,
follows from the measurement of the quantity
on the left-hand side of
Eq.~(\ref{deltaEratio2}).
Recent atomic measurements can provide bounds at the level
of accuracy
$10^{-15}\mbox{yr}^{-1}$,
as stated above.

\par
In view of
Eq.~(\ref{deltaEratio2})
and assuming that the value
$\Delta E (\alpha, Z)$
almost tends to zero at some value of $Z$
(crossing-point), we find that the coefficient
\begin{eqnarray}
\eta
&=&\label{defeta}
(\Delta E)^{-1}
\alpha
\,
\frac{\partial \Delta E(\alpha, Z)}{\partial \alpha}
\end{eqnarray}
plays the role of an enhancement factor
when extracting the value for
$\dot{\alpha} / \alpha$
provided the left-hand side of
Eq.~(\ref{deltaEratio2})
is known.
Indeed, differentiating
$\Delta E (\alpha, Z)$
we find
\begin{eqnarray}
\Delta E (\alpha, Z_0)
&=&\label{deltaEeq0}
0
\,,
\\
\alpha
\frac{\partial \Delta E (\alpha, Z_0)}{\partial \alpha}
&\ne&\label{deltaEeq0next}
0
\,,
\end{eqnarray}
where
$Z_0$ (in general, a real number)
would correspond to the exact crossing point.

\par
Note, that for ions the
equality~(\ref{deltaEeq0})
holds only approximately, since the nuclear charge number $Z$
takes integer values, only.
Nevertheless, the cancellation of different terms in
the expression for
$\Delta E (\alpha, Z_0)$
can reduce its value by several orders of magnitude,
while such a cancellation does not occur in
Eq.~(\ref{deltaEeq0next}).
For $Z = 66$
the energy difference
$\Delta E_{   2{}^1\!S_0  -  2{}^3\!P_0  }(\alpha,Z)$
in two-electron ions reduces to
$0.016 \mbox{\rm eV}$,
while the individual terms in
Eq.~(\ref{deltaEf})
contributing to
Eq.~(\ref{deltaEeq0next})
are of the order 
$10^2 \mbox{\rm eV}$.
Thus, we can expect a value for the enhancement factor
$\eta$
of the order
$10^3$ (see
\cite{andreev03}).

\par
It should be emphasized,
that in order to make use of this almost
level crossing,
one has to be able to measure directly the energy difference
$\Delta E (\alpha, Z)$,
otherwise the gain of the enhancement factor
$10^3$
will be lost by the requirement to measure the positions
of each of the crossing levels with an accuracy better by
$3$
order of magnitude.
Fortunately, such direct measurements are feasible when applying
laser techniques for the spectroscopy of HCI.
Energy differences of the order of
$1 \mbox{\rm eV} - 0.1 \mbox{\rm eV} $
correspond to the optical or infrared frequency intervals.
A number of experiments with lasers has already been performed
for measurements of hyperfine-structure intervals in HCI,
which lay in the optical region
\cite{klaft94,seelig98}.

\par
To become sensitive to temporal variations of
$\dot{\alpha} / \alpha$
at the level of
$10^{-15} \mbox{\rm yr}^{-1} $
in laser experiments with HCI performed during a time interval
$\delta t$ about $1 \mbox{\rm yr}$,
it will require a relative accuracy in the measurement
of the energy difference of the order
$10^{-12}$.
Here we have taken into account the enhancement factor
$10^3$.
In principle, this accuracy could be achieved in laser resonant
measurements.
In particular, the two-photon transition for
$1s$-$2s$
states in hydrogen has been measured with a relative accuracy
of about
$10^{-14}$
\cite{niering00}.
Such a level of precision has not yet been
reached in experiments with HCI.
Nevertheless, as a precondition of any experiments
theory should provide most accurate values for
$\Delta E (\alpha, Z)$
although
the goal to achieve an experimental
accuracy of
$10^{-12}$
is still far beyond present abilities.
In laser resonance measurements in neutral atoms
the theoretical accuracy is much poorer than the experimental one.
This is a usual situation, which
should not be considered as a hindrance
for the search of a temporal variation of
$\alpha$ in HCI.

\par
Let us consider in more detail the crossing of the levels
$2{}^1\! S_0$
and
$2{}^3\! P_0$
in two-electron ion with nuclear charge number
$Z=66$.
Since the one-photon
$0-0$
transition is forbidden, one should
either investigate
transitions between the hyperfine sublevels
in the isotopes with the nonzero nuclear spin
or look for
the two photon
$E1M1$
resonance.
The
$E1M1$
transitions
$2{}^3\! P_0 - 1{}^1\! S_0$
in two-electron HCI were studied in
Ref.~\cite{drake85,savukov02,labzowsky04}.
We should stress that the width of the mentioned resonance
is quite large due to the $E1E1$ transition
$2 {}^1\! S_0$ -
$1 {}^1\! S_0$:
$\Gamma ( 2 {}^1 S_0 \to  2\gamma (E1) + 1 {}^1\! S_0)
\simeq 1\times 10^{12} \mbox{\rm s}^{-1}
= 4 \times 10^{-3} \mbox{\rm eV}$
\cite{derevianko97}
what is only 4 times smaller than our value for the
transition frequency
$2 {}^1\! S_0 - 2 {}^3 \!P_0$
for
$Z=66$.

\par
For each of the corrections
Eqs.~(\ref{deltaEf1ph}-\ref{deltaEfscrsevp})
the corresponding values for the coefficients
$C^{\text{correction}}_{\text{powers of}\, \alpha, Z}$
in the vicinity of
$Z=66$,
can be deduced via interpolation.
Continuous curves
$\Delta E^{\text{correction}}(Z)$
are generated employing the results of the calculations
presented in
\cite{andreev03}.
The adjustment for the value
$Z=66$
yields the following results (in  units of eV)
\begin{eqnarray}
\Delta E^{\text{NS}}
&=&\label{adjustmentcoeff}
1.869
\,,
\\\nonumber
C^{\text{1ph}}_{0,1}&=&0.162\,,\,\,
C^{\text{1ph}}_{2,3}=-1.259\,,\,\,
C^{\text{1ph}}_{4,5}=-1.691\,,
\\\nonumber
C^{\text{2ph}}_{0,0}&=&-3.341\,,\,\,
C^{\text{2ph}}_{2,2}=19.489\,,\,\,
C^{\text{2ph}}_{4,4}=-31.669\,,
\\\nonumber
C^{\text{SE+VP}}_{3,4}&=&1.803\,,\,\,
C^{\text{Scr(SE+VP)}}_{3,3}=-0.761
\,.
\end{eqnarray}
The relatively large values for the coefficients
$C^{\text{2ph}}$
indicate the poor convergence of the series expansion
(\ref{deltaEf2ph}).
This does not mean, however, the inaccuracy in the result.
One has to remember that the coefficients in
(\ref{adjustmentcoeff}) are
obtained from the numerical all-order in $\alpha Z$ calculation
and the poor convergence of $\alpha Z$ expansion for any
correction in the left-hand side of
Eq.~(\ref{adjustmentcoeff})
does not influence
our conclusions.
The fit of the exact numerical function by
the few $\alpha Z$ expansion terms for the fixed value of $Z$
(in our case $Z=66$) can be done as accurate as necessary and
only the large values of the coefficients for $C^{\text{2ph}}$
correction compared to expansion coefficients to other corrections
are puzzling.
The minimum value for the total energy shift at
$Z = 66$
turns out to be
$\Delta E(\alpha, Z) = -0.016 \mbox{\rm eV}$.
Calculation of the derivative according to
Eqs.~(\ref{deltaEf}-\ref{deltaEfscrsevp})
leads to
\begin{eqnarray}
\alpha
\frac{\partial \Delta E (\alpha, Z)}{\partial \alpha}
&=&\label{logder}
  2C^{\text{1ph}}_{2,3}\alpha^2 Z^3
+ 4C^{\text{1ph}}_{4,5}\alpha^4 Z^5
+ 2C^{\text{2ph}}_{2,2}\alpha^2 Z^2
+ 4C^{\text{2ph}}_{4,4}\alpha^4 Z^4
\\\nonumber
&&
+ 3C^{\text{SE+VP}}_{3,4}\alpha^3 Z^4
+ 3C^{\text{Scr(SE+VP)}}_{3,3}\alpha^3 Z^3
\,,
\end{eqnarray}
respectively after
insertion of the
coefficients~(\ref{adjustmentcoeff})
$
\alpha
({\partial \Delta E (\alpha, Z)}/{\partial \alpha})
=
-2.1 \times 10^{1}
\mbox{\rm eV}
$.
Hence the enhancement factor
$\eta$
at $Z=66$ is obtained as
$\eta = 1.3 \times 10^{3}$.
It was found that
the inaccuracy of
the employed scheme of interpolation
is less than
$6 \%$.

\par
The feasibility of the proposed experiment
depends crucially on whether
tunable lasers within the frequency range of about
$0.01 - 0.02 \mbox{\rm eV}$,
respectively,
with wavelengths between
$60 - 120 \mbox{$\mu$\rm m}$
are available.
In particular,
the Free Electron Lasers (FELs)
satisfy these conditions,
though they do not provide
the required relative accuracy
$10^{-12}$.
In particular, the Forschungzentrum Rossendorf (Dresden)
provides
ELBE FELs
\cite{fahmy03}
with wavelengths
$\lambda=3-25 \mbox{$\mu$\rm m}$
and it is supposed
to reach the region
$\lambda=20-150 \mbox{$\mu$\rm m}$.
Alternative methods may exploit
relativistic Doppler tuning,
multiphoton resonances and optical mixing techniques.
E.g. CO$_{2}$
lasers have typical wavelengths
in the region of about
$10.6 \mbox{\rm $\mu$m}$.
Consider an ion with
relativistic factor
$\gamma = 1/\sqrt{1 - \beta^{2}}$
colliding head-on
with a photon of frequency $\omega_{\mbox{\rm lab}}$
in the laboratory frame.
In the ions rest frame the frequency
$\omega$
is given by
$\omega =\gamma(1+\beta)\omega_{\mbox{\rm lab}}$.
The recent GSI ESR
facility provides
$\gamma=1.048$ and
up to
$\gamma=23$ at
the new projected GSI storage ring
\cite{gsi01}.
Manipulating between these two values one can
cover the range of wavelengths
$10.6 \mbox{\rm $\mu$m} \le \lambda \le 488
\mbox{\rm $\mu$m}$.
The same concerns tunable semiconductor-diode laser PbSnTe with
wavelength
$\lambda_{\mbox{\rm lab}}$
within the range
$6 - 30 \mbox{\rm $\mu$m}$.
Furthermore, one could think of exciting the levels
under consideration via
absorption of $n$ photons of frequency
$\omega$
such that
$\Delta E= n \omega$ (multiphoton resonance)
\cite{huber99,eikema01}.
Finally, we mention that
the optical mixing techniques
\cite{demtroeder98}
can also help to achieve
radiation sources with wavelengths within the range
$1 - 100 \mbox{\rm $\mu$m}$.

\par
The most severe problem which arises when performing
the proposed experiment is how to achieve the required accuracy
$10^{-12}$ in case of the broad resonances.
As mentioned above
the width of the
$2 {}^1\! S_0$ -$2 {}^3\! P_0$
resonance in a He-like ion with $Z=66$ is about
$10^{12} \mbox{\rm s}^{-1}$.
Accordingly, the problem consists in the observation
of time-dependence derivation of natural line profile.
The latter is characterized by the maximum frequency
$\omega_{\mbox{\rm max}}$ and the
width $\Gamma$.
Although the value $\omega_{\mbox{\rm max}}$ may not coincide
with $\Delta E$
\cite{labzowsky01prl}
it exhibits the same enhancement
behaviour due temporal variation of
$\alpha$.
Thus, one has to concentrate on
the measurement of $\omega_{\mbox{\rm max}}$,
which,
in principle, appears as a question of statistics.
Suppose that
the measurement occurs in the vicinity of
$\omega_{\mbox{\rm max}}$ within the
laser bandwidth that is assumed to be much smaller than $\Gamma$.
Then each photon absorption can be considered as statistical ``event'',
or elementary act of measurement.
An accuracy $r=10^{-12}$ implies
to register $r^{-2} = 10^{24}$ ``events''.

\par
The required intensity of the laser beam may be estimated
according to the formula
$r^{-2} = T N_{\mbox{\rm ion}} \sigma I_{\mbox{\rm ph}}$.
Here $T$ denotes the time duration of the laser experiment
(e.g. even $T = 3\times10^{7} \mbox{s}$
may be still realistic),
$I_{\mbox{\rm ph}}$
is the photon flux
(in units photons/s)
and
$N_{\mbox{\rm ion}}$ is the number of ions, meeting the laser beam per second.
The latter value may be estimated according to
$N_{\mbox{\rm ion}} = N_{\mbox{\rm tot}} N_{\mbox{\rm rev}}$,
where
$N_{\mbox{\rm tot}} = 10^{8}/ 0.01 \, \mbox{cm}^{2}$
is the total number of the ions in the storage ring
(assuming a beam diameter of about $0.1 \mbox{\rm cm}$)
and
$N_{\mbox{\rm rev}}$ is the number of
revolutions per second.
A rough estimate for the photon-absorption cross-section
$\sigma$ is provided by
$\sigma \simeq (a_{0})^{2}/Z^{2}$,
where $a_{0}$ is the Bohr
radius.
For $Z=66$
this yields about
$\sigma \simeq 10^{-20} \mbox{\rm cm}^{2}$.
Taking all together one obtains
the necessary laser beam intensity of the
order
$I_{\mbox{\rm ph}} \simeq 3\times 10^{20} \mbox{\rm photon/s}$.
Although
this is a very huge number one should
keep in mind first,
that the future GSI facilities
will provide an ion beam being
more intensive by several orders of magnitude and
second, that the progress in laser techniques has been
very fast during the last years.

\par
Very stringent, in principle, is also the problem with Doppler
broadening.
The latter is defined by the expression
$\Gamma_{D}=\gamma\beta(\Delta v/v)\omega_{\mbox{\rm lab}}$.
While the natural broadening is time-translational invariant
the Doppler
broadening depends on the beam temperature, which can
also vary during the time period $T$.
Fortunately, the Doppler broadening appears to be
smaller than the natural one in case of
$2 {}^1\! S_0- 2 {}^3\! P_0$
transition.
According to
\cite{stohlker00}
the relative velocity spread
in the beam is given by
$\Delta v/v \simeq 5\times10^{-4}$,
which yields
$\Gamma_{D} \simeq 2\times 10^{-4} \omega_{\mbox{\rm lab}}$, i.e.
much smaller than $\Gamma$.
Once the problem with the natural
broadening will be solved, the Doppler broadening problem will
be solved as well.

\par
It should be mentioned,
that similar atomic experiments utilizing levels crossing
with both HCI and neutral atoms
in the presence of
an external magnetic or electric field can also be proposed.
However, the reliability of such experiments will strongly
depend on
the constancy of the strength of the external fields during
the time interval
$\delta t \approx 1 \mbox{\rm yr}$.
In our case the ``constancy'' of the
$Z$
value is assured by the charge conservation law.

\par
We deeply regret the sudden decease of our friend and coauthor
Prof. Dr. Gerhard Soff
before completion of this investigation.

\par
The authors are grateful to
D.~Glazov, A.~Prozorov and V.~Sharipov and A.~Volotka
for the fruitful discussion.
This work was supported in parts by the INTAS grant
No~0354-3604.
G.P. and G.S. acknowledge financial support
from BMBF, DFG and GSI.
The work of O.Y.A. and L.N.L. has been supported in parts
by the DFG,
by the RFBR Grant No. 02-02-16578
and by Minobrazovanie grant No.~E02-3.1-7.
L.N.L. is grateful
to GSI (Darmstadt) and
to the TU Dresden for the
hospitality during his visits in 2004.



\begin{thebibliography}{10}
\providecommand*{\bibinfo}[2]{#2}
\providecommand*{\eprint}[1]{#1}
\providecommand*{\url}[1]{#1}
\bibitem{dirac37}
\bibinfo{author}{P.~A.~M. Dirac}, \bibinfo{journal}{Nature ({\it London})}
  \bibinfo{volume}{\textbf{139}}, \bibinfo{pages}{323} (\bibinfo{date}{1937}).
\bibitem{uzan03}
\bibinfo{author}{\mbox{J.-P}. Uzan}, \bibinfo{journal}{Rev. Mod. Phys.}
  \bibinfo{volume}{\textbf{75}}, \bibinfo{pages}{403} (\bibinfo{date}{2003}).
\bibitem{webb01}
\bibinfo{author}{J.~K. Webb}, \emph{et~al.},
  \bibinfo{journal}{Phys. Rev. Lett.} \bibinfo{volume}{\textbf{87}},
  \bibinfo{pages}{091301} (\bibinfo{date}{2001}).
\bibitem{fujii02}
\bibinfo{author}{Y.~Fujii}, \emph{et~al.}, \bibinfo{journal}{Int. J. Mod. Phys. D}
  \bibinfo{volume}{\textbf{11}}, \bibinfo{pages}{1137} (\bibinfo{date}{2002}).
\bibitem{fischer04}
\bibinfo{author}{M.~Fischer}, \emph{et~al.},
  \bibinfo{journal}{Phys. Rev. Lett} \bibinfo{volume}{\textbf{92}},
  \bibinfo{pages}{230802} (\bibinfo{date}{2004}).
\bibitem{peik04}
\bibinfo{author}{E.~Peik}, \emph{et~al.},
  \bibinfo{journal}{Phys. Rev. Lett} \bibinfo{volume}{\textbf{93}},
  \bibinfo{pages}{170801} (\bibinfo{date}{2004}).
\bibitem{andreev03}
\bibinfo{author}{O.~\mbox{Yu}. Andreev}, \emph{et~al.},
  \bibinfo{journal}{Phys. Rev. A} \bibinfo{volume}{\textbf{67}},
  \bibinfo{pages}{012503} (\bibinfo{date}{2003}).
\bibitem{klaft94}
\bibinfo{author}{I.~Klaft}, \emph{et~al.},
\bibinfo{journal}{Phys. Rev. Lett.}
  \bibinfo{volume}{\textbf{73}}, \bibinfo{pages}{2425} (\bibinfo{date}{1994}).
\bibitem{seelig98}
\bibinfo{author}{P.~Seelig}, \emph{et~al.},
  \bibinfo{journal}{Phys. Rev. Lett.} \bibinfo{volume}{\textbf{81}},
  \bibinfo{pages}{4824} (\bibinfo{date}{1998}).
\bibitem{niering00}
\bibinfo{author}{M.~Niering}, \emph{et~al.},
  \bibinfo{journal}{Phys. Rev. Lett.} \bibinfo{volume}{\textbf{84}},
  \bibinfo{pages}{5496} (\bibinfo{date}{2000}).
\bibitem{drake85}
\bibinfo{author}{G.~W.~F. Drake}, \bibinfo{journal}{Nucl. Instr. Meth.}
  \bibinfo{volume}{\textbf{B9}}, \bibinfo{pages}{465} (\bibinfo{date}{1985}).
\bibitem{savukov02}
\bibinfo{author}{I.~N. Savukov} and \bibinfo{author}{W.~R. Johnson},
  \bibinfo{journal}{Phys. Rev. A} \bibinfo{volume}{\textbf{66}},
  \bibinfo{pages}{062507} (\bibinfo{date}{2002}).
\bibitem{labzowsky04}
\bibinfo{author}{L.~N. Labzowsky} and \bibinfo{author}{A.~V. Shonin},
  \bibinfo{journal}{Phys. Rev. A} \bibinfo{volume}{\textbf{69}},
  \bibinfo{pages}{012503} (\bibinfo{date}{2004}).
\bibitem{derevianko97}
\bibinfo{author}{A.~Derevianko} and \bibinfo{author}{W.~R. Johnson},
  \bibinfo{journal}{Phys. Rev. A} \bibinfo{volume}{\textbf{56}},
  \bibinfo{pages}{1288} (\bibinfo{date}{1997}).
\bibitem{fahmy03}
\bibinfo{author}{K.~Fahmy}, \emph{et~al.},
  \bibinfo{journal}{J. Bio. Phys.} \bibinfo{volume}{\textbf{29}},
  \bibinfo{pages}{303} (\bibinfo{date}{2003}).
\bibitem{gsi01}
\bibinfo{author}{{GSI, Conceptual Disign Report for the International
  Accelerator Facility for Beams of Ions and Antiprotons (2001)}}.
\bibitem{huber99}
\bibinfo{author}{A.~Huber}, \emph{et~al.},
  \bibinfo{journal}{Phys. Rev. A} \bibinfo{volume}{\textbf{59}},
  \bibinfo{pages}{1844} (\bibinfo{date}{1999}).
\bibitem{eikema01}
\bibinfo{author}{K.~S.~E. Eikema}, \emph{et~al.},
 \bibinfo{journal}{Phys. Rev. Lett.}
  \bibinfo{volume}{\textbf{86}}, \bibinfo{pages}{5679} (\bibinfo{date}{2001}).
\bibitem{demtroeder98}
\bibinfo{author}{W.~Demitr{\"o}der}, \bibinfo{title}{\emph{Laser Spectroscopy}}
  (\bibinfo{publisher}{Springer-Verlag}, \bibinfo{year}{1998}).
\bibitem{labzowsky01prl}
\bibinfo{author}{L.~N. Labzowsky}, \emph{et~al.},
  \bibinfo{journal}{Phys. Rev. Lett.} \bibinfo{volume}{\textbf{87}},
  \bibinfo{pages}{143003} (\bibinfo{date}{2001}).
\bibitem{stohlker00}
\bibinfo{author}{\mbox{Th}. St{\"o}hlker}, \emph{et~al.},
  \bibinfo{journal}{Phys. Rev. Lett.} \bibinfo{volume}{\textbf{85}},
  \bibinfo{pages}{3109} (\bibinfo{date}{2000}).

\end{thebibliography}

\end{document}